\documentclass[aps,prb,showpacs,tightenlines,twocolumn,nofootinbib,nobibnotes,superscriptaddress]{revtex4-1}
\usepackage{amsmath,amssymb,amsfonts,bm}
\usepackage{graphicx}
\usepackage{epstopdf}
\usepackage{dcolumn}
\usepackage{mathrsfs}
\usepackage[colorlinks=true,linkcolor=blue,citecolor=blue, urlcolor=blue,bookmarks=false]{hyperref}

\usepackage{natbib}

\begin{document}

\title{Floquet topological insulator phase in a Weyl-semimetal thin film with disorder}
\date{\today }

\author{Rui Chen}

\author{Dong-Hui Xu}

\author{Bin Zhou}
\thanks{binzhou@hubu.edu.cn}

\affiliation{Department of Physics, Hubei University, Wuhan 430062, China}

\begin{abstract}
We investigate the effects of periodic fields and disorder on topological properties of a Weyl-semimetal thin film. The two periodic fields, i.e., a periodic magnetic field and elliptically polarized light, are discussed respectively. By use of the Floquet theory, we find that both the two periodic drives can resonantly induce the topological transitions from normal insulator (NI) phases to Floquet topological insulator (FTI) phases. The Floquet topological transitions are characterized by variation of Chern number. Moreover, we show that the Floquet topological transitions can be explained by a combination of the quantum well approximation and the rotating wave approximation. In the disordered Weyl-semimetal thin film model under periodic fields, we calculate the Bott index to characterize topological phase. It is found that the FTI phase is robust against weak disorder, and collapses for strong disorder strength. Interestingly, we find that disorder can also induce a topological transition from a topological trivial phase to an FTI phase, establishing the Floquet topological Anderson insulator (FTAI) phase. Finally, an effective-medium theory based on the Born approximation further confirms the numerical conclusions.
\end{abstract}

\maketitle

\section{Introduction}
In the last decade, topological phases have attracted an enormous amount of attention in condensed matter community\cite{Hasan2010RMP,Qi2011RMP}. A typical example is the quantum anomalous Hall insulator (QAHI), which is interesting as it has topologically protected chiral edge states in the bulk band gap, and supports the quantum Hall effect without Landau levels\cite{Weng2015AdvPhys,Liu2016AnnualReview}. Due to its topologically nontrivial properties and enormous potential applications for designing novel spintronics devices, there have been tremendous developments and interests in searching for a robust QAHI. The QAHI was first proposed by Haldane in a hexagonal lattice model with a periodic flux\cite{Haldane1988PRL}. Subsequently, the QAHI was also predicted in several related systems, such as disorder-induced Anderson insulators\cite{Onoda2003PRL}, Rashba spin-orbit coupled lattices\cite{Qiao2010PRB,Chen2017PLA}, optical lattices\cite{Wu2008PRL,Zhang2011PRB}, and magnetic topological insulators\cite{Wang2015PhysScr,Liu2008PRL,Jiang2012PRB,Yu2010Science}. In the meantime, significant progresses have been achieved in experiments. The QAHI has been observed experimentally in magnetically doped topological insulators\cite{Chang2013Science,Chang2015NatMat}.

Time-periodic drive provides a new method to control electronic states, and can even achieve topologically nontrivial phases in topologically trivial systems\cite{Wang2013Science,Mahmood2016NatPhys,Sie2014NatPhys,Kim2014Science}. The topologically nontrivial phase induced by time-periodic drive is referred to as Floquet topological insulator (FTI) phase, which has also been proposed in various systems, including periodically driven semiconductors \cite{Lindner2011NatPhys,Lindner2013PRB}, Rashba nanowires\cite{Klinovaja16PRL,Thakurathi17PRB}, graphenelike systems \cite{Kitagawa2011PRB,Oka2009PRB,Delplace2013PRB,GomezLeon2014PRB} and other two-dimensional (2D) systems\cite{Mikami2016PRB}. Comparing with its solid-state counterpart, the FTI phase in the context of artificial time-periodic systems, is more easily achievable. For example, the Haldane model has been realized in a periodically modulated optical honeycomb lattice by use of ultra-cold fermionic atoms\cite{Jotzu2014Nature}, and an optical system composed of an array of evanescently coupled helical waveguides with the propagation coordinate $z$ playing the role of time\cite{Rechtsman2013Nature}. In addition, the Floquet topological phases in three-dimensional materials, such as three-dimensional topological insulators\cite{Calvo2015PRB,Fregoso2013PRB}, Weyl semimetals\cite{Bucciantini2017PRB}, Dirac semimetals\cite{Wang2014EPL,Chan2016PRL,Ebihara2016PRB}, and line-node semimetals\cite{Chan2016PRB,YanZ2016PRL,Narayan2016PRB,Taguchi2016PRB1,Chen2018PRBLNSM}, have been studied extensively.

In time-independent system, disorder-induced topological phase, referred to as topological Anderson insulator (TAI), has been well studied in a variety of systems\cite{LiJ2009PRL,Guo2010PRL,Jiang2009PRB,Li2011PRB,Zheng2018arXiv,Wu2016CPB,Su2016PRB,Chen2017PRBDirac,Chen2017PRBLieb,Song2012PRB,Liu2016PRL}. In the Born approximation, the phase transition can be explained that disorder renormalizes the topological mass which determines the topological phase of the system\cite{Groth2009PRL}. Very recently, the observation of a TAI in one-dimensional disordered atomic wires has been reported experimentally\cite{Meier2018Science}. Interestingly, in time-periodic systems, it is also reported that disorder can renormalize either the system parameters or the form of the time-periodic drives, then inducing the phase transitions from the normal insulator (NI) to the Floquet topological Anderson insulator (FTAI)\cite{Titum2017PRB,Titum2015PRL,Titum2016PRX,Roy2016PRB,Yap2017PRB,Shtanko2018PRL,Nathan2017arXiv,Kundu2017arXiv,Du2018PRB}. Furthermore, owing to the rapid development of topological photonics\cite{Lu2014NaturePhotonics}, the FTAI has been observed experimentally in a photonic platform recently\cite{Stutzer2018Nature}.

Weyl semimetals, a newly discovered class of gapless topological phases, are characterized by linearly dispersive
band-touching points called Weyl nodes\cite{Wan2011PRB}, and exhibit exotic transport properties
such as the chiral anomaly induced negative magnetoresistance\cite{Xiong2015Science,Huang2015PRX,Zhang2016NatCom} and the Weyl orbit physics\cite{Potter2014NatCom,Moll2016Nature,Dai2016NatPhys,Zhang2017NatPhys}. Owing to the quantum confinement effect, the spectrum of a Weyl-semimetal thin film will split into subbands and an energy gap is opened at the Weyl nodes\cite{Xiao2015SciRep,Pan2015SciRep,Takane2016JPSJ}. The system becomes a quasi-2D NI when the film is thin enough. In this paper, we adopt the Floquet theory to investigate the effects of a periodic magnetic field and an elliptically polarized light on a Weyl-semimetal thin film. It is found that both the two periodic drives can resonantly induce the exotic FTI phases. While the Chern number characterizing the FTI phase depends on the driving type, frequency and amplitude, which can be understood by a combination of quantum well approximation and the rotating wave approximation. Then we focus on the effects of disorder on the periodically driven systems. In the disordered Weyl-semimetal thin film model under periodic fields, by computing the Bott index, we find that disorder can induce a topological transition from a topologically trivial phase to an FTAI phase. The numerical results are confirmed by an effective-medium theory based on the Born approximation.

This paper is organized as follows: In Sec.~\ref{model},
we first introduce a Hamiltonian describing a Weyl-semimetal thin film, the quantum well approximation, and the Floquet theory. Then, we study the effects of a periodic magnetic field and an elliptically polarized light on a Weyl-semimetal thin film, respectively, in Sec.~\ref{periodicdrive}. Next, we focus on the effect of disorder in a periodically driven system in Sec.~\ref{Disorder}. Finally, a brief summary is presented in Sec.~\ref{Conclusion}.

\section{model}
\label{model}
\subsection{Weyl-semimetal thin film}
\label{thinfilm}

Our starting point is a simple two-band tight-binding model describing a Weyl semimetal discretized on a cubic lattice, which is given by the following Hamiltonian\cite{Yang2011PRB},
\begin{align}
H_0\left( \mathbf{k}\right)   =&\left(  m_{z}-t_{z}\cos k_{z}\right)
\sigma_{z}+m_{0}\left(  2-\cos k_{x}-\cos k_{y}\right)  \sigma_{z}\nonumber\\
&  +t_{x}\sigma_{x}\sin k_{x}+t_{y}\sigma_{y}\sin k_{y}, \label{Hk}%
\end{align}
where $t_{i}(i=x,y,z)$, $m_{z}$, and $m_{0}$ are the model parameters. $\sigma_{i}$ are Pauli matrices representing spin. This model breaks the time-reversal symmetry, and can be regarded as a layered 2D QAHI coupled to its neighboring layers by the spin dependent tunneling $t_z$. In the following calculations, we take the unity lattice constant $a=1$ and fix $t_x=t_y=t_z=t$, $m_0/t=1$, and $m_z/t=0.8$. The condition $| m_{z}/m_{0}|< t_{z}/m_{0}$ guarantees that the system depicts a Weyl semimetal phase hosting a pair of Weyl nodes locating at $\left( 0, 0, \pm k_z^0 \right) $ with $k^0_z=\arccos{(m_{z}/t_{z})}$\cite{Yang2011PRB,Chen2015PRL,Chen2018PRBWeyl}.

\begin{figure}[ptb]
\includegraphics[width=8cm]{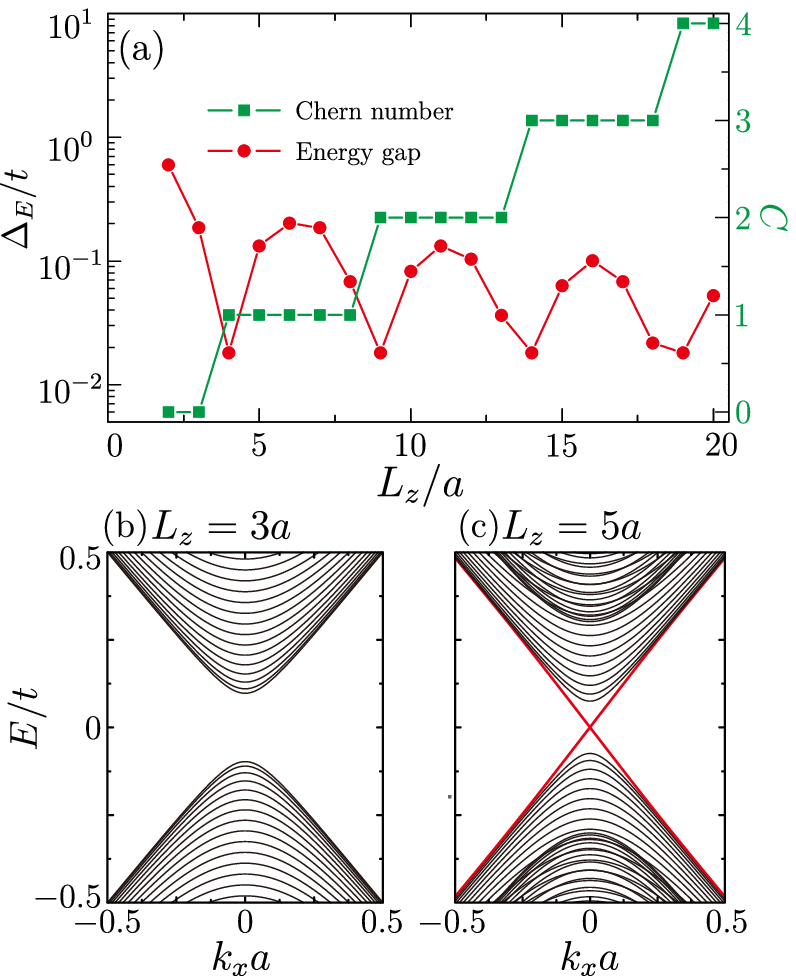}
\caption{(Color online) (a) The Chern number (the green square line) and the confinement-induced energy gap (the red circle line) as functions of the thin-film thickness $L_z$. (b) and (c) are band structures for Weyl-semimetal thin film with open boundary conditions in the $y$ and $z$ directions and periodic boundary condition in the $x$ direction. The system size is chosen as $L_y=100a$, and the thin-film thickness is set as $L_z=3a$ in (b) and $L_z=5a$ in (c). The black lines are the bulk states, and the red lines correspond to the chiral edge states on the side surfaces.}
\label{fig1}
\end{figure}

We would like to consider a Weyl-semimetal thin film configuration, with confinement along $z$ direction and thin-film thickness $L_z(=n_z a)$. As a direct consequence of quantum confinement effect\cite{Xiao2015SciRep,Pan2015SciRep}, the bulk spectrum of the quasi-2D Hamiltonian $H_0\left(\mathbf{k}_\Vert, n_z  \right)$ will be quantized into discrete subbands along $z$ direction. In Fig.~\ref{fig1}(a), the confinement induced band gap $\Delta_{E}/t$ and the total Chern number $C$ of the bands below the band gap are plotted as functions of the thin-film thickness $L_z$. It is found that the band gap oscillatingly decays with the increasing of $n_z$, accompanied by the variation of Chern number at each dip.

By expanding $H_0\left( \mathbf{k}\right)$ at $\mathbf{k}=\left(0,0,0\right)$ and using quantum well approximation\cite{Xiao2015SciRep,Pan2015SciRep}, this phenomenon can be well explained through the effective Hamiltonian $H_{\text{eff}}\left(\mathbf{k}_\Vert, n  \right)$ for each subband $n\left(=1,2,\cdots,n_z\right)$, which takes the form
\begin{equation}
H_{\text{eff}}\left(\mathbf{k}_\Vert, n  \right)   =\mathbf{d}_n\cdot\bm{\sigma},
\end{equation}
where $\mathbf{d}_n=\left(t_{x}k_{x},t_{y}k_{y},M_n-m_0k_\Vert^2/2\right)$, $\bm{\sigma}=\left(\sigma_x,\sigma_y,\sigma_z\right)$ and $M_n=m_z-t_z\left[1-\left(n\pi/L_z\right)^2/2\right]$
is the mass term determining the topological properties of these subbands. For each subband $n$, the above effective Hamiltonian is equivalent to a single $2\times 2$ block of the Bernevig-Hughes-Zhang (BHZ) model\cite{Bernevig2006Science} Hamiltonian describing the 2D topological insulator, and its Chern number is defined as $C_{\text{eff}}^n=\left[\text{sign}{\left(M_n\right)}+
\text{sign}{\left(m_0\right)}\right]/2$\cite{Lu2010PRB}. Therefore, the total Chern number of the effective Hamiltonians is given by $C_{\text{eff}}=\sum_n C_{\text{eff}}^n$, which is obviously determined by the thin-film thickness $L_z$.

In general, for a system with an open boundary condition, Chern number $C$ characterizes the number of the edge states. When the film is thin enough $\left(n_z \leq 3\right)$, such that all the subbands are topologically trivial, the system is insulating and topologically trivial indicated by the Chern number $C=0$ [Fig.~\ref{fig1}(b)]. On changing the thin-film thickness, the conduction and valence bands cross, it is possible to tune the model from a topologically trivial phase to a topologically nontrivial phase (i.e., QAHI phase), with the emergence of a pair of chiral edge states characterized by a nonzero Chern number $C=1$ (e.g. $n_{z}=5$) [Fig.~\ref{fig1}(c)]. As $n_{z}$ further increases, more pairs of chiral edge states would appear in the band gap of the system, characterized by higher Chern numbers with $C=2,3,\cdots$.

\subsection{Floquet Hamiltonian}

Now we introduce the Floquet theory\cite{Titum2017PRB} for a time-periodic Hamiltonian, $H\left( t \right)=H\left( t +T\right)$ with the period $T=2\pi/\Omega$, and $\Omega$ is the frequency. By employing the Floquet theory, the wave function of the time-periodic Schr\"odinger equation $ i\partial_t \Psi\left(t\right)=
 H\left(t \right)\Psi\left(t\right)$, has the form $\Psi\left(t\right)=\sum_{m}\psi_m e^{-i\left(\epsilon+m\Omega\right)t}$, where $\epsilon$ is called the quasienergy and the summation over $m$ takes all integer values. After a Fourier series expansion, it is found that solving the time-periodic Schr\"odinger equation is actually equivalent to solving an eigenvalue problem of a static system, $\sum_{m} H^F_{nm}\psi_m=\epsilon\psi_n$, where \begin{equation}
 H^F_{nm}=n\Omega\delta_{nm}+\frac{1}{T}\int_0^T dt H\left(t \right)e^{i\Omega \left(n-m\right)t},
 \label{Floquet}
 \end{equation}
is a block Hamiltonian of the Floquet state. $n$ and $m$ are integers ranging from $-\infty$ to $\infty$, but we only need to truncate the Hamiltonian $H^F$ into finite dimensions with $-n_F<m,n<n_F$ in numerical calculation, where $n_F$ is determined by use of convergence tests. Moreover, if $\Psi\left(t\right)$ is an eigenvector with quasienergy $\epsilon$, then $e^{in\Omega t}\Psi\left(t\right)$ should also be an eigenvector of the system, but with a quasienergy $\epsilon+n\Omega$. Therefore, all the Floquet states can be restricted in a quasienergy scale of $\Omega$, which is the so-called Floquet zone.

\begin{figure}[ptb]
\includegraphics[width=8cm]{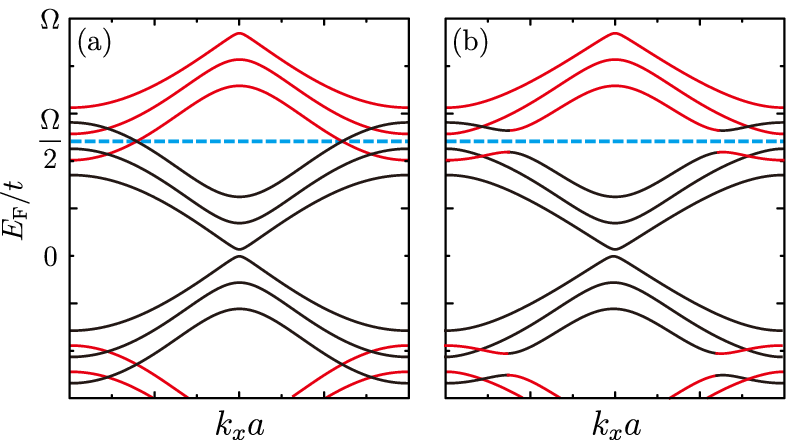}
\caption{(Color online) (a) Schematic illustrations of the band spectrum obtained from the diagonal blocks of the Floquet Hamiltonian $H^{\text{F}}$. The black and red lines correspond to the spectrum of the undriven and resonant blocks, and they cross with each other at $E_F/t=\Omega/2$ (the blue dashed line). (b) The energy spectrum of the full Floquet Hamiltonian. The off-diagonal blocks of $H^{\text{F}}$ hybridize the quasienergies shown in (a) and gap them out. Here, the thin-film thickness is set as $L_z=3a$, and $k_y=0$. }
\label{fig2}
\end{figure}

In the subsequent content, we will restrict our attention to the resonant Floquet zone ranging from $0<\epsilon<\Omega$, where the conduction and valence bands resonantly couple with each other. To explain this, as shown in Fig. \ref{fig2}(a), we plot the quasienergies of a Floquet Hamiltonian $H^{\text{F}}$, where we only consider the diagonal blocks and set the off-diagonal blocks to be zero. Actually, each diagonal block is just a copy of the original block $H_{00}^{\text{F}}$, shifting in energy by $n\Omega$. When the driving frequency satisfies $\max{\left[\Delta_E,W/2\right]}<\Omega<W$, where $\Delta_E$ and $W$ are the values of the band-gap and band-width of the original block, the conduction and valence bands would mix with each other through a resonance. The off-diagonal blocks, representing transitions between the diagonal blocks via photons absorption or emission processes, will hybridize the resonant quasienergies and gap them out [Fig. \ref{fig2}(b)]. Next, we will show a Floquet topological phase can be induced by a periodically driven field, even if the system is in a topologically trivial phase (i.e., $n_z=3$).

\section{periodic drive}
\label{periodicdrive}

The time-periodic part Hamiltonian, $H\left(t\right)$, can be experimentally induced either from a periodic magnetic field, or elliptically polarized light\cite{Lindner2011NatPhys,Titum2017PRB}. In this section, we will investigate them separately.

\subsection{Periodic magnetic field}
\label{Zeemanfield}
The time-dependent Hamiltonian can be introduced through a periodic Zeeman coupling term in the presence of a periodically modulated magnetic field as
\begin{equation}
H_{\text{Zee}}\left(t \right)=H_0\left(\mathbf{k}_\Vert, n_z  \right)+I_{n_z}\otimes\mathbf{V}\cdot\bm{\sigma} \cos \left(\Omega t\right) ,
\end{equation}
where $\mathbf{V}=\left(V_x,V_y,V_z\right)$. In experiments, the periodic Zeeman field can be realized by a microwave-terahertz (THz) oscillating magnetic field\cite{Lindner2011NatPhys,Kampfrath11NatPho,Kampfrath13NatPho}. By use of the previous results in Eq. \ref{Floquet}, the time-periodic Hamiltonian $H_{\text{Zee}}\left(t\right)$, describing a Weyl-semimetal thin film with a periodic Zeeman coupling effect, can be explicitly represented in Floquet matrix form $H^F_{\text{Zee}}$.

\begin{figure}[ptb]
\includegraphics[width=8cm]{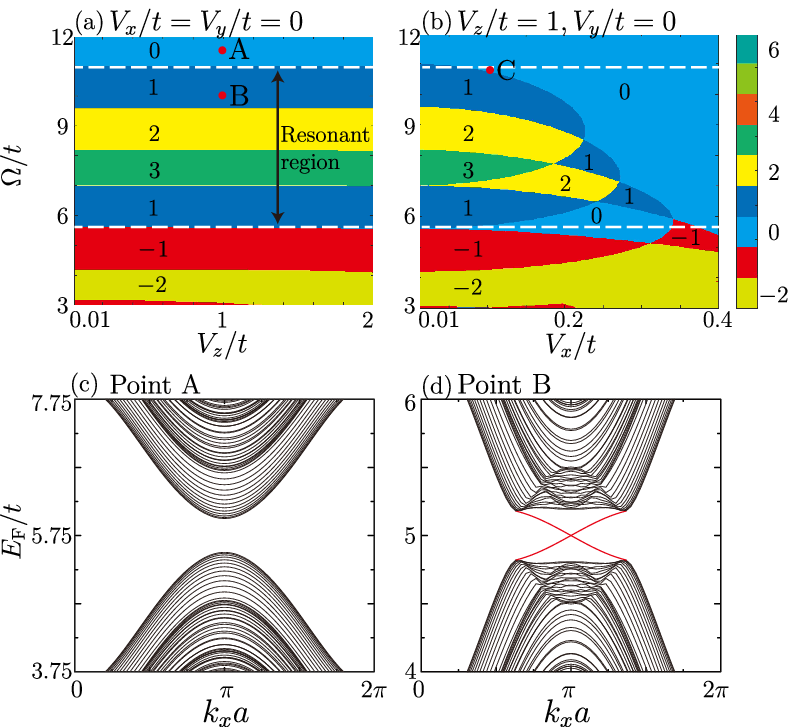}
\caption{(Color online) (a) and (b) are phase diagrams for the periodic magnetic field driven Weyl-semimetal thin film in the $\left(\Omega/t,V_z/t\right)$ plane with $V_x/t=V_y/t=0$, and the $\left(\Omega/t,V_x/t\right)$ plane with $V_z/t=1$ and $V_y/t=0$, respectively. The color map corresponds to value of the Chern number. The region which lies between the white dashed lines corresponds to resonant region satisfying $\max{\left[\Delta_E,W/2\right]}<\Omega<W$. (c) and (d) are the quasienergy spectra of the Floquet Hamiltonian with $V_z/t=1$ and $V_x/t=V_y/t=0$, which correspond to the point A $\left(\Omega/t=11.5\right)$ and point B $\left(\Omega/t=10\right)$ labeled in (a), respectively. The black lines are the bulk states, and the red lines correspond to the chiral edge states. }
\label{fig3}
\end{figure}

To characterize the system, through the numerical method proposed by Fukui \emph{et al}\cite{Fukui2005JPSP}, we calculate the corresponding Chern number of the bands under the resonant energy $E_F=\Omega/2$ for $H^F_{\text{Zee}}$ with a truncation of the Floquet matrix for $n_F=3$. We will first study the phase diagram of the system with $V_x=V_y=0$. As shown in Fig.~\ref{fig3}(a), we plot the Chern number as functions of the $z$-direction Zeeman field amplitude $V_z$ and the driving frequency $\Omega$. When the frequency $\Omega$ is higher than the band width $W/t=11$, it is found that the conduction bands and the resonant valence bands are well separated, and the corresponding Chern number is zero. However, when the frequency gets lower, the lowest resonant valence subband touches with the highest conduction subband, resulting in a band inversion and opening a topologically nontrivial gap with the Chern number $C=1$. Further decreasing $\Omega$, more resonant valence subbands would gradually cross with the conduction subbands, giving rise to higher Chern numbers $C=2$ and $C=3$. When the frequency is lower than the band width of the conduction subbands $(W/2t\approx5.5)$, due to the multiple folding of the Floquet bands with many band touchings, the phase diagram becomes quite complicated, where we will not concentrate in this paper\cite{Mikami2016PRB}. Thus, we show that the nontrivial FTI can be induced even in a topologically trivial phase of Weyl-semimetal thin film. In addition, numerical results show that a weak periodic Zeeman field can gap out the Weyl-semimetal thin film, and the non-trivial FTI phase can be induced by a weak periodic magnetic field. Therefore, in order to simplify numerical calculations, one omits the orbital effect of a periodic magnetic field, which does not change the results qualitatively.

To examine our results, we plot the energy band spectra for the Weyl-semimetal thin film under open boundary conditions along $y$ and $z$ directions with different frequencies in Figs.~\ref{fig3}(c) and \ref{fig3}(d), which correspond to the points A and B in Fig.~\ref{fig3}(a). For the case of $\Omega/t=11.5$ in Fig.~\ref{fig3}(c), the conduction bands are well separated with the resonant valence bands, and the gap is topologically trivial. For the case of $\Omega/t=10$ in Fig.~\ref{fig3}(d), the conduction bands hybridize with the resonant valence bands, and the gapless topological edge states connecting the conduction bands and the resonant valence bands can be easily distinguished. For a lower frequency, more pairs of chiral edge states will arise in the resonant gap.  Therefore, the appearance of the chiral edge states further confirms the nontrivial topology of the driven system. It should be mentioned that the precise correspondence between the Chern number and the number of the edge states does not always work in 2D Floquet systems\cite{Rudner2013PRX}, but we find that the Chern number is still valid to characterize the Floquet topological phase in the Weyl-semimetal thin-film model by checking the bulk-edge correspondence.

Now we introduce $V_x$ to the present model, and the phase diagram is shown in Fig.~\ref{fig3}(b) with $V_z/t=1$ and $V_y/t=0$. The results show that a nonzero $V_x$ tends to destroy the topologically nontrivial phase induced by $V_z$ in the resonant region $\max{\left[\Delta_E,W/2\right]}<\Omega<W$. Moreover, it is found that $V_x$ and $V_y$ play the same role in modulating the phase diagrams.

We have mentioned in Sec.~\ref{thinfilm} that a Weyl-semimetal thin film can be regarded as a superposition of several BHZ models. By combining the quantum well approximation\cite{Xiao2015SciRep,Pan2015SciRep} and the rotating wave approximation\cite{
Lindner2011NatPhys,Titum2017PRB}, the band topology in the resonantly driven Weyl-semimetal thin film can also be understood by a superposition of several effective Hamiltonians with
\begin{equation}
H_{\text{eff}}^{F}\left(\mathbf{k}_\Vert, n  \right)=\frac{\Omega}{2} I_2+\left(\left|\mathbf{d}_n\right|-\frac{\Omega}{2} \right)\hat{d}_n\cdot\bm{\sigma}+\frac{1}{2}\mathbf{T}_n\cdot\bm{\sigma},
\label{RWA}
\end{equation}
where
\begin{equation}
\mathbf{T}_n=\mathbf{V}-\left(\mathbf{V}\cdot\hat{d}_n\right)\hat{d}_n\nonumber,
\end{equation}
and $\hat{d}_n=\mathbf{d}_n/\left|\mathbf{d}_n\right|$ is a unit vector. For a nonzero Zeeman field, it is found that each effective Hamiltonian opens a gap at the resonant energy, $E_{\text{eff}}^F=\Omega/2$. For $V_x/t=V_y/t=0$ and $V_z/t \neq0$, the inverted gap is found to be topologically nontrivial with the Chern number $C_\text{eff}^F=1$, while for $V_z/t =0$ and $V_x/t,V_y/t \neq0$, the gap is found to be trivial with $C_\text{eff}^F=0$. Therefore, the topological properties of the system depend on the competition effect of $V_z$ and $V_x$ ($V_y$).

\subsection{Elliptically polarized light}

The time-dependent Hamiltonian can also be introduced by normally illuminating with elliptically polarized light described by a time-varying gauge field $\mathbf{A}\left(t\right)=\left(A_x \sin\Omega t,A_y \cos\Omega t\right)$, where $A_x$ and $A_y$ determine the ellipticity of the incident light. Elliptically polarized light turns into circularly polarized light when $A_x=A_y$, and linearly polarized light when $A_x=0$ or $A_y=0$. By use of the Peierls substitution, the time-dependent Hamiltonian is obtained as
\begin{equation}
H_{\text{Lig}}\left(t\right)=H_0\left(\mathbf{k}_\Vert-\mathbf{A}\left(t\right), n_z  \right).
\end{equation}

\begin{figure}[ttb]
\includegraphics[width=8cm]{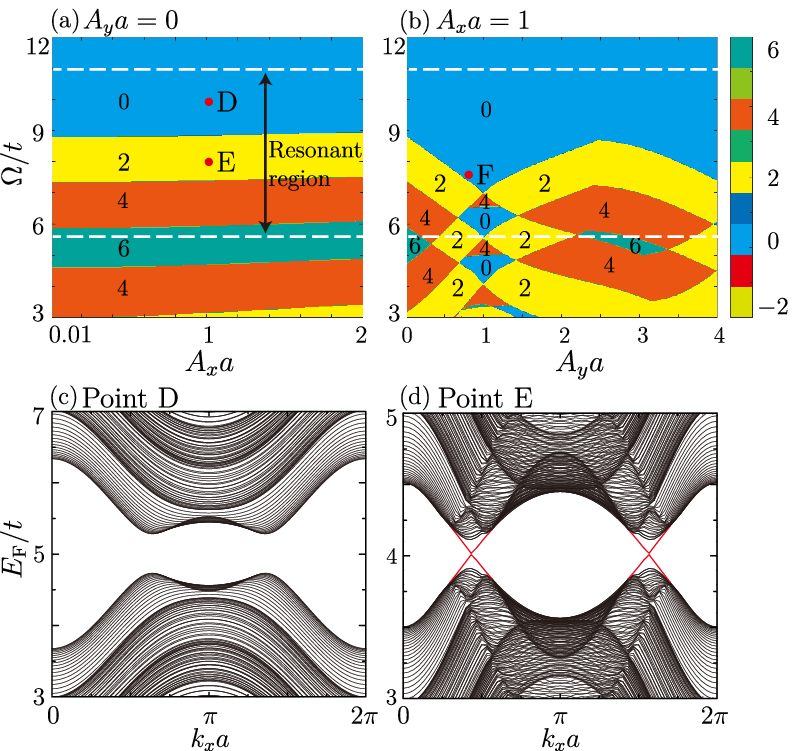}
\caption{(Color online) (a) and (b) are phase diagrams for the polarized light driven Weyl-semimetal thin film in the $\left(\Omega/t,A_x a\right)$ plane with $A_y a=0$, and the $\left(\Omega/t,A_y a\right)$ plane with $A_x a=1$, respectively. The color map corresponds to value of the Chern number. The region which lies between the white dashed lines corresponds to resonant region satisfying $\max{\left[\Delta_E,W/2\right]}<\Omega<W$. (c) and (d) are the quasienergy spectra of the Floquet Hamiltonian with $A_x a=1$ and $A_y a=0$, which correspond to the point D $\left(\Omega/t=10\right)$ and point E $\left(\Omega/t=8\right)$ labeled in (a), respectively. The black lines are the bulk states, and the red lines correspond to the chiral edge states.
  }
\label{fig4}
\end{figure}

We will first consider linearly polarized light, i.e., $A_y=0$. Using the same method in Sec.~\ref{Zeemanfield}, we obtain the phase diagram for the Chern number as a function of the light frequency $\Omega/t$ and the $x$-direction light amplitude $A_x a$ in Fig.~\ref{fig4}(a). The results are similar to the previous case that the system is subjected to a periodic magnetic field, in which various Floquet topological phases appear in the resonant frequency. But there are still differences in the two cases. In the resonant region, the Chern number takes positive integers $C=1,2,\cdots,n_z$ for the case of a periodic magnetic field, while it takes even numbers $C=0,2,4,\cdots,2 n_z$ for the case of linearly polarized light.

The band spectra under open boundary conditions along $y$ and $z$ directions are shown in Figs.~\ref{fig4}(c) and \ref{fig4}(d), which correspond to the points D and E in Fig.~\ref{fig4}(a), with the frequencies $\Omega/t=10$ and $\Omega/t=8$, respectively. For the case of $\Omega/t=10$, in which the conduction subbands hybridize with the resonant valence subbands, the system is still insulating with a topologically trivial gap characterized by the Chern number $C=0$ [Fig.~\ref{fig4}(c)]. This is different from the case under a periodic magnetic field, that the FTI phase with nonzero Chern number $C=1$, occurs immediately when these subbands hybridize [Fig.~\ref{fig3}(d)]. For a lower frequency $\Omega/t=8$ shown in Fig.~\ref{fig4}(d), there are two pair of chiral edge states, corresponding to the Chern number $C=2$.

We also plot the phase diagram for the system illuminated by elliptically polarized light with $A_x=1$ in Fig.~\ref{fig4}(b). The results show that the ellipticity of the polarized light can obviously influence the topological properties of the system. For circularly polarized light with $A_x=A_y$, depending on the frequency, the Chern number of the system can only be $C=0$ or $C=4$.

By use of the rotating wave approximation, the effective Hamiltonian describing the Weyl-semimetal thin film under elliptically polarized light has the same form as the Hamiltonian (\ref{RWA}), except that $\mathbf{T}_n$ is given by
\begin{align}
\mathbf{T}_n=&T_n^R\left(-\cos\theta_n\cos\phi_n,
-\cos\theta_n\sin\phi_n,\sin\theta_n\right)\nonumber
\\
&+T_n^I\left(-\sin\phi_n,\cos\phi_n,0\right),
\end{align}
where
\begin{align}
T_n^R=&\frac{t_x A_x}{2}\sin\phi_n\cos k_x+\frac{t_y A_y}{2}\cos\theta_n\sin\phi_n\cos k_y\nonumber
\\
&+\frac{m_0A_y}{2}
\sin\theta_n\sin k_y\nonumber,
\\
T_n^I=&-\frac{t_x A_x}{2}\cos\theta_n\cos\phi_n\cos k_x-\frac{t_y A_y}{2}\cos\phi_n\cos k_y\nonumber
\\
&-\frac{m_0A_x}{2}
\sin\theta_n\sin k_x\nonumber,
\end{align}
and we have used the spherical coordinate, $\mathbf{d}_n=\left|\mathbf{d}_n\right|\left(\sin\theta_n
\cos\phi_n,\sin\theta_n\sin\phi_n,\cos\theta_n\right)$.

For each effective Hamiltonian, the inclusion of light also opens a resonant gap at $E_{\text{eff}}^F=\Omega/2$. For linearly polarized light, the Chern number $C_\text{eff}^F$ of the inverted bands under the resonant gap is $2$, while for circularly polarized light, the Chern number is found to be $0$. Thus, the ellipticity of the incident polarized light influences the topological phase of the driven system.

\section{Disorder}
\label{Disorder}

Generally, in a QAHI the Chern number $C$ always connects with the Hall conductance whose value is quantized as $\sigma_{xy}=Ce^2/h$\cite{Thouless1982PRL,Hatsugai1993PRL}. However, recent studies have shown the Hall response might be different for Floquet systems\cite{Kundu2014PRL,Torres2014PRL}. For example, the Hall conductance in the Floquet state of a circularly polarized light irradiated graphene deviates from integer values at some Floquet topological transitions\cite{Kundu2014PRL}. Remarkably, this suppressed conductance can be dramatically enhanced by disorder, which means disorder may play an important role in the FTI system.

In this section, we discuss the effect of disorder on the Floquet topological phases induced by a periodic magnetic field and elliptically polarized light by computing the Bott index and using the Born Approximation. We will introduce the on-site disorder through random on-site energy with a uniform distribution within $\left[-U/2,U/2\right]$, where $\pm U/2$ define the largest and smallest disorder amplitude on one site.

\subsection{Bott index}
We will use the Bott index\cite{Loring2010EPL} to characterize the disordered time-periodically driven system. For a time-independent Hamiltonian with disorder, the Bott index is equivalent to the real space Chern number, which determines the Hall conductivity of the system\cite{Hastings2010JMP}. Recently, the Bott index is generalized to obtain the Chern number of a disordered periodically driven system, by adopting a truncated Floquet Hamiltonian $H^F$ defined on real space with periodic boundary conditions\cite{Rudner2013PRX}. Considering two diagonal matrices $X_{i,i}=x_i$ and $Y_{i,i}=y_i$, where $\left(x_i,y_i\right)$ are coordinates of the $i$-th lattice, one can obtain the two unitary matrices $U_X=\exp\left(i2\pi X/L_x\right)$ and $U_Y=\exp\left(i2\pi Y/L_y\right)$, where $L_x$ and $L_y$ are the system size. Using the eigenstates of the bands below the resonant gap, one obtains the projector $P$, and the projected unitary matrices with $\tilde{U}_{X,Y}=PU_{X,Y}P$. For every random disorder configuration, the Bott index $C_b^F=(1/2\pi)\text{Im}\left[\text{Tr}\left(
\text{ln}\tilde{U}_Y\tilde{U}_X\tilde{U}_Y^\dag
\tilde{U}_X^\dag\right) \right]$ is an integer, and the averaged Bott index is computed by averaging various disorder configurations until it converges. In this section, we take the truncation of the Floquet matrix for $n_F=2$, as the calculation for the Bott index is computationally very demanding. Moreover, we find that the real space Bott index obtained in the clean limit is in accordance with the $k$-space Chern number obtained previously.

\begin{figure}[ptb]
\includegraphics[width=8cm]{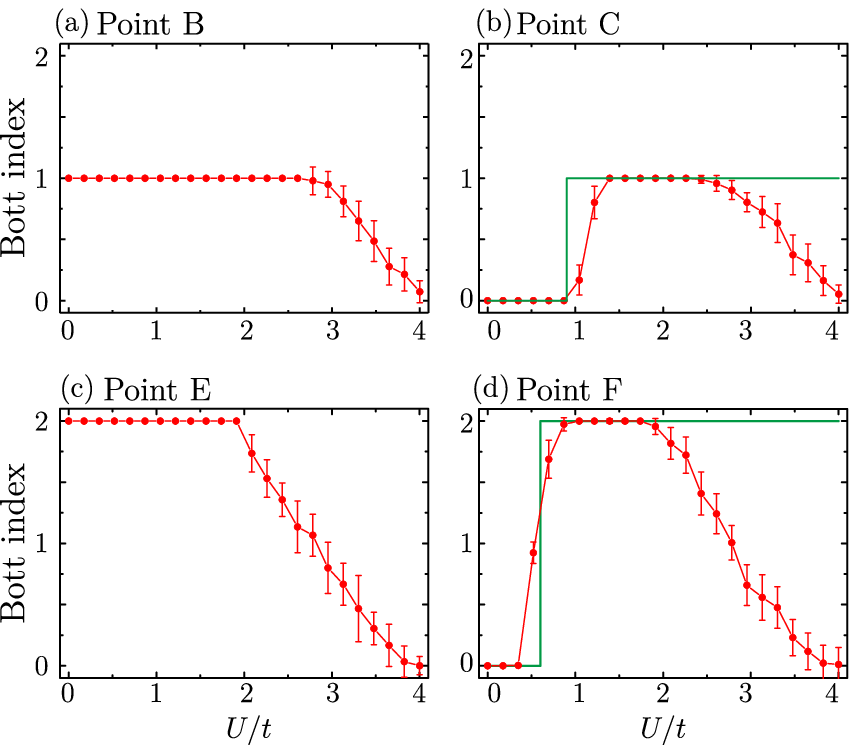}
\caption{(Color online) The Bott index varies as a function of the disorder strength $U$. (a), (b), (d), and (d) correspond to the points B, C, E, and F shown in Fig.~\ref{fig3} and Fig.~\ref{fig4}, respectively. The error bars show the standard deviation of the Bott index for 500 samples. The green lines in (b) and (d) are the Chern numbers of the renormalized Floquet Hamiltonian obtained by the Born approximation.}
\label{fig5}
\end{figure}

Firstly, we consider the case of the system driven by a periodic magnetic field, and plot the Bott index versus the disorder strength $U$ in Figs.~\ref{fig5}(a) and \ref{fig5}(b), which correspond to points B and C in Figs.~\ref{fig3}(a) and \ref{fig3}(b). When the system is in the topologically nontrivial regime [Fig.~\ref{fig5}(a)], the Bott index is quantized to be $1$ and quite stable for weak disorders. As expected, with increasing disorder strength, the quantized plateau drops and eventually decreases to zero, indicates a phase transition from a topological insulator to a Anderson insulator. This result shows the FTI phase in a Weyl-semimetal thin film is robust again weak disorder. Interestingly, as shown in Fig.~\ref{fig5}(b), we find disorder can also trigger an adverse topological transition. In the clean limit, the Bott index is zero as the system is topologically trivial. However, with increasing disorder strength, the Bott index becomes finite and reaches a quantized value $1$. The plateau maintains for a certain range of disorder strength and eventually decreases. The disappearance of the fluctuation of the Bott index indicates that the quantized value is originated from the topologically protected edge states. Therefore, we identify this disorder and drive induced phase as a FTAI phase.

We have also investigated the case of the system driven by elliptically polarized light, and the Bott index varies as a function of the disorder strength $U$ is shown in Figs.~\ref{fig5}(c) and \ref{fig5}(d), which correspond to points E and F in Figs.~\ref{fig4}(a) and \ref{fig4}(b). The findings are similar to the case of the system driven by a periodic magnetic field, that disorder can induce the transitions from topologically nontrivial phases to trivial phases [Fig.~\ref{fig5}(c)] or vice verse [Fig.~\ref{fig5}(d)], except that the value of the quantized plateau which characterizes the topologically nontrivial phase is $2$.

\subsection{Born approximation}
In order to gain more insights into these phase transitions at weak disorder, we analyze the truncated Floquet Hamiltonian by use of the Born approximation\cite{Groth2009PRL,Titum2015PRL}. In the zero-order Born approximation, the self-energy is given by
\begin{equation}
\Sigma_{\text{Zee(Lig)}}=\frac{U^2}{12}\left(\frac{a}{2\pi}\right)^2\int_{\text{FBZ}}d \mathbf{k}_\Vert\frac{1}{E_f-H^F_{\text{Zee(Lig)}}\left(\mathbf{k}_\Vert \right)},
\end{equation}
where $H^F_{\text{Zee(Lig)}}\left(\mathbf{k}_\Vert \right)$ is the truncated Floquet Hamiltonian of the system driven by a periodic magnetic field (or elliptically polarized light). This approximation is valid when the magnitude of $\Sigma$ is much smaller than $E_f$\cite{Shon1998JPSP,Koshino2006PRB}. The
coefficient $1/12$ originates from the variance $\left\langle U^{2}\right\rangle
=U^{2}/12$ of a random variable uniformly distributed in the range $\left[
-U/2,U/2\right] $. This integration is
over the first Brillouin zone (FBZ). In the numerical calculation, we fix the Fermi level $E_f=\Omega/2$ as we concentrate on the resonant quasienergy gap. By calculating the Chern number of the renormalized Floquet Hamiltonian $\tilde{H}^F_{\text{Zee(Lig)}}\left(\mathbf{k}_\Vert \right)=H^F_{\text{Zee(Lig)}}\left(\mathbf{k}_\Vert \right)+\Sigma_{\text{Zee(Lig)}}$ as functions of $U$, we obtain the curves shown in Figs.~\ref{fig5}(b) and \ref{fig5}(d). It is found that the results based on the Born approximation match well with the numerical calculations in the weak disorder region. The disorder has a renormalization effect on the quasienergies of the truncated Floquet Hamiltonian, leading to the exotic FTAI phases.

\section{Conclusion}
\label{Conclusion}

In this paper, we firstly investigate the Floquet topological transitions in a Weyl-semimetal thin film under the periodically driven fields. The effects of a periodic magnetic field and elliptically polarized light are discussed, respectively. The topological phases are characterized by the Chern number. The computation of the Chern number indicates that both the two periodic drives can resonantly induce the topological transitions from NI phases to FTI phases. The Floquet topological transitions can be explained by a combination of the quantum well approximation and the rotating wave approximation. Next, we investigate the effect of disorder in a periodically driven Weyl-semimetal thin film. Based on the numerical results of the Bott index, it is shown that the FTI phase is robust against weak disorder, while strong disorder will lead to Anderson localization and induce topological transitions from FTI phases to NI phases. The computation of the Bott index also reveals another role of disorder, and we observe a transition from a trivial insulating phase to an FTI phase at a finite disorder strength. Moreover, the result obtained by the effective-medium theory based on the Born approximation also confirms the occurring of the Floquet topological Anderson transition in a Weyl-semimetal thin film.

Though the studies above mentioned are concentrated on the Weyl-semimetal thin films, we believe the FTI and FTAI phases can also occur in the Dirac-semimetal thin films, as the Dirac semimetal can be regarded as two copies of Weyl semimetal with different chiralities. For a Dirac-semimetal thin film, the system crosses over between normal insulators and quantum spin Hall insulator, and respects the time-reversal symmetry\cite{Xiao2015SciRep,Pan2015SciRep}. Therefore, we suppose the interplay between periodic drive, disorder and topology on a Dirac-semimetal thin film would produce more new exotic quantum phenomena.

We expect the FTI and FTAI phases can be experimentally realized in the TaAs family of Weyl semimetals\cite{Xu2015Science}, and the Dirac semimetal materials Cd$_3$As$_2$\cite{Wang2013PRB} and Na$_3$Bi\cite{Wang2012PRB}. For example, for a Dirac semimetal Na$_3$Bi with thickness $n_z=2.5\text{ nm}$, the finite size gap is about $50\text{ meV}$\cite{Xiao2015SciRep,Wang2012PRB,Chen2017PRBDirac}. In our theoretical calculations, we choose high driven frequencies to avoid the interferences from the multiple-resonance effect as we concentrate on the single-resonance process. While in experiments, the frequency is just required to be larger than the finite size gap $\Delta_E$ of the undriven system, as the multiple-resonance effects will be highly suppressed by adjusting the driven intensities and frequencies\cite{Lindner2011NatPhys}. To realize the light driven FTI phase, we choose the photon energy to be $\hbar \Omega\approx100\text{ meV}$, the amplitude of linearly polarized light to be $A_L\approx0.02 \text{ \AA}^{-1}$. The corresponding electric field strength $E=\hbar\Omega A_L/e$ is about $2\times 10^7 \text{ V/m}$, which is within experimental accessibility\cite{Wang2013Science,Mahmood2016NatPhys}. In addition, the periodic Zeeman field can be experimentally realized by a THz oscillating magnetic field. The frequency of the periodic electromagnetic-wave radiation reported ranges from $1$ THz to $10$ THz\cite{Kampfrath11NatPho,Kampfrath13NatPho}, and the corresponding energy is from about $4$ meV to $40$ meV. On the other hand, the finite size gap is highly controllable by varying the thin-film thickness\cite{Xiao2015SciRep} or using a vertical electric field\cite{Pan2015SciRep}. For example, the finite size gap in a Dirac semimetal Na$_{3}$Bi is about $15$ meV when the thickness is $n_z=5$ nm, and about $5$ meV when $n_z=10$ nm\cite{Xiao2015SciRep}. Thus, all of these features offer the possibility of realizing our proposal on a Weyl-semimetal thin film.

\section*{Acknowledgments}

B.Z. was supported by the National Natural Science Foundation of China (Grant No. 11274102), the Program for New Century Excellent Talents in University of Ministry of Education of China (Grant No. NCET-11-0960), and the Specialized Research Fund for the Doctoral Program of Higher Education of China (Grant No.
20134208110001). R.C. and D.-H.X. were supported by the National Natural Science Foundation of China (Grant No. 11704106) and the Scientific Research Project of Education Department of Hubei Province (Grant No. Q20171005). D.-H.X. also acknowledges the support of the Chutian Scholars Program in Hubei Province.

\end{document}